\documentclass[twocolumn,showpacs,preprintnumbers,superscriptaddress,amsmath,amssymb,floatfix,aps]{revtex4}

\usepackage[dvips]{graphicx}
\usepackage{dcolumn}
\usepackage{bm}
\begin{document}

\title{Transient nanobubbles in short-time electrolysis }

\author{Vitaly B. Svetovoy}
\email[Corresponding author: ]{v.svetovoy@utwente.nl}
\affiliation{MESA$^+$ Institute for Nanotechnology, University of
Twente, PO 217, 7500 AE Enschede, The Netherlands}
\affiliation{Institute of Physics and Technology, Yaroslavl Branch,
Russian Academy of Sciences, 150007, Yaroslavl, Russia}
\author{Remco G. P. Sanders}
\affiliation{MESA$^+$ Institute for Nanotechnology, University of
Twente, PO 217, 7500 AE Enschede, The Netherlands}
\author{Miko C. Elwenspoek}
\affiliation{MESA$^+$ Institute for Nanotechnology, University of
Twente, PO 217, 7500 AE Enschede, The Netherlands}
\affiliation{FRIAS, University of Freiburg, 79104 Freiburg, Germany}

\date{\today}

\begin{abstract}

Water electrolysis in a microsystem is observed and analyzed on a
short-time scale $\sim 10\;\mu$s. Very unusual properties of the
process are stressed. An extremely high current density is observed
because the process is not limited by the diffusion of electroactive
species. The high current is accompanied by a high relative
supersaturation $S>1000$ that results in homogeneous nucleation of
bubbles. On the short-time scale only nanobubbles can be formed.
These nanobubbles densely cover the electrodes and aggregate at a
later time to microbubbles. The effect is significantly intensified
with a small increase of temperature. Application of alternating
polarity voltage pulses produces bubbles containing a mixture of
hydrogen and oxygen. Spontaneous reaction between gases is observed
for stoichiometric bubbles with the size smallaer than $150\;$nm.
Such bubbles disintegrate violently affecting the surface of
electrodes.

\end{abstract}
\pacs{47.55.D-, 64.60.qj, 82.33.Vx}

\maketitle

\section{Introduction}\label{intro}

More than 200 years gone since people observed water electrolysis for
the first time (see \cite{Tra99,Lev99} and references therein). Since
then the process is widely used in different applications including
hydrogen production \cite{Kha98,Zen10}. Nowadays electrolysis of
water find also applications in different kind of microdevices such
as actuators \cite{Nea96,Cam02,Hue02,Lee10}, pumps
\cite{Ate04,Men08}, and others \cite{Lee05,Wan07}. Fast performance
of microsystems requires processing on a short-time scale.
Surprisingly enough, most of the researches on water electrolysis
dealt with macro or microscopic systems on a time scale longer than
milliseconds \cite{Ate04,Mat09}. A few papers, where shorter
(microseconds) voltage pulses were considered, discussed increase of
the overall efficiency \cite{Shi06} or optimization of the
performance of electrolysis \cite{Wei03}, but did not observe the
process on the microsecond time scale.

In normal long-time electrolysis (long-time means here the time scale
$\tau > 1$ ms) gas bubbles with the diameters in the range
$10-1000\;\mu$m are observed \cite{Mat09,Bra85}. The maximal current
density that can be reached in this case is around 1
A$\cdot$cm$^{-2}$ \cite{Vog05} and the maximal relative
supersaturation $\sim 100$ \cite{Vog93}. These values cannot be
increased further because bubbles cover all the surface of the
electrode making it inactive \cite{Vog05,Hin75,Vog83}.

In the short-time electrolysis, say $\tau=1-100\;\mu$s, one could
expect unusual properties of the process. It becomes obvious if we
consider gas diffusion into the liquid above the electrode. For the
time $\tau$ a significant supersaturation is possible at the distance
from the electrode $l\sim\sqrt{D\tau}$, where $D$ is the diffusion
coefficient of the formed gas in the surrounding liquid. Taking as
typical values $D\sim 10^{-9}\;$m$^2$/s and $\tau\sim 10\;\mu$s one
finds $l\sim 100\;$nm. It means that on this timescale only
nanobubbles can be formed.This is indeed the case as we demonstrated
recently \cite{Sve11}. Nanobubbles of $200-300\;$nm in diameter were
observed stroboscopically above microelectrodes. These nanobubbles
were formed for the first $20-50\;\mu$s then aggregated to
microbubbles after $100-300\;\mu$s, but the following dynamics was
very slow.

A number of observations indicate that smaller nanobubbles can exist
at shorter times $\tau\lesssim 10\;\mu$s. For example, it was
demonstrated that the reaction between hydrogen and oxygen can be
ignited spontaneously in nanobubbles with the size smaller than
$150\;$nm \cite{Sve11}. Different manifestations of this reaction
show that the reaction persists for nanobubbles as small as $50\;$nm
or even smaller.

In this paper we describe formation of nanobubble on the short-time
scale $\tau=1-100\;\mu$s. It is demonstrated that very high current
densities and supersaturations can exist accompanied by the
homogeneous nucleation of nanobubbles. The bubbles containing
hydrogen or oxygen grow and aggregate forming micro-sized bubbles on
the timescale $\tau\gtrsim100\;\mu$s. The nanobubbles containing
mixture of the gases disappear as the result of the reaction between
gases. These are transient nanobubbles, which exist only a short time
and difficult for direct observation. In contrast with the first
report \cite{Sve11} we provide here detailed description of the
transient nanobubbles and pay less attention to the chemical reaction
inside of these bubbles.

The nanobubbles discussed in this paper differ from the surface
nanobubble, which attracted recently a considerable attention. The
long-lived surface nanobubbles were observed with different methods
\cite{Lou00,Ste03,Swi04}. Observation with atomic force microscopes
(AFM) became especially popular (see \cite{Wan10} and references
therein) due to its simple realization. These nanobubbles are widely
discussed in the literature because of their mysterious stability.
High Laplace pressure drives diffusion of gas out of the bubble
resulting in a dissolution time of $\sim 10\;\mu$s but not days as
observed (see \cite{Sed12} for a review). Precise reason for this
stability is not clear yet but in a promising model \cite{Bre08} of
"dynamical equilibrium" the gas flux going out of the bubble through
the spherical cap is balanced with a gaseous influx at the contact
line.

A number of methods to generate the surface nanobubbles are in use
including solvent exchange, temperature change for liquid or
substrate, and liquid pressurizing \cite{Yan07}. The electrochemical
formation of the surface nanobubbles were observed in water
electrolysis for hydrogen \cite{Zha06} and for both hydrogen and
oxygen \cite{Yan09} on highly orientated pyrolytic graphite. The
electrochemical process is considered as a flexible way to control
the bubble production (density and size) \cite{Yan11}. In contrast
with the electrochemical process considered in this paper the
electrochemical production of the surface nanobubbles had much
smaller current density ($\sim 10^{-4}\;$A/cm$^{2}$) and much longer
time scale ($\sim 10\;$s).

This paper is organized as follows. In Sec. \ref{system} we describe
the system for observation of the short-time electrolysis,
application of the voltage pulses, and the current response. In Sec.
\ref{strob} the main results obtained with the stroboscopic
observations are described. Electrolysis with alternating sign pulses
is described in Sec. \ref{alternating} where we present the results
obtained with a stroboscop and vibrometer. We explain also
modification of the electrode surface due to the process. In Sec.
\ref{discussion} the evolution of the transient nanobubbles is
summarized and we discuss their possible applications. Our
conclusions are collected in the last section.

\section{Short-time electrolysis}\label{system}

\begin{figure}[tb]
  \includegraphics[width=8cm]{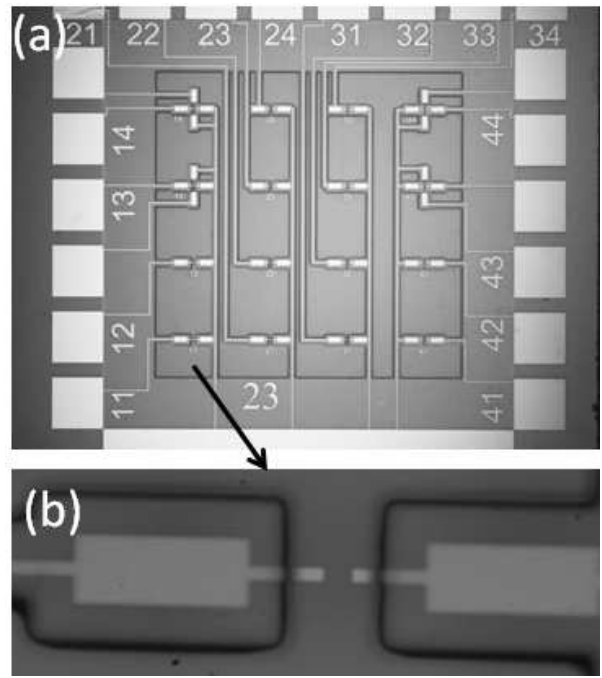}
  \caption{(a) Microchip used for investigation of
  electrolysis on short-time scales. (b) Pair of microelectrodes
  that can be addressed separately.
  }\label{fig1}
  \vspace{-0.5cm}
\end{figure}

To analyze electrolysis on a short-time scale in a microscopic system
special microchips were prepared. One of the chips is shown in Fig.
\ref{fig1}. For fabrication we used wafers made of borofloat glass
BF-33. First, Ti sublayer was deposited on glass ($10\;$nm) for
better adhesion, then the 100 nm thick metallic layer was sputtered.
We used the wafers with different metals on top such as Pt, Pd, Au,
or W. The metals were patterned and covered lithographically with the
insulating resist SU8 where it was necessary. Different thicknesses
of SU8 layer of 3 and $90\;\mu$m were in use. The fabricated wafers
were diced on separate chips of size $12\times 12\;$mm$^2$. Each chip
contained 16 pairs of the microelectrodes of different shapes and
sizes and the contact pads to address a specific pair electrically.
The patterned SU8 layer formed walls of open channels for a liquid. A
chip filled in with the electrolyte was covered by a thing
($30\;\mu$m) glass plate for observation of the electrolysis.

As the solution for electrolysis we used $15\;$g of Na$_2$SO$_4$
dissolved in $100\;$ml of deionized water. Also we have used the
solutions of NaCl and KI in water in similar concentrations.

The process was observed with a homemade stroboscopic system
\cite{Bro08}. In this system the light source and the electrical
pulses applied to the electrodes were controlled with the time
resolution $0.1\;\mu$s. As a light source two powerful green LEDs
LUXEON$^{\circledR}$ III were used. With this system we have been
able to observe the electrodes in an optical microscope with the
illumination time longer than $5\;\mu$s.

\subsection{Voltage}\label{volt}

For a chosen pair of the electrodes one was always grounded and the
other one (working electrode) could be at a negative or positive
potential. Hydrogen or oxygen bubbles were produced above the working
electrode. We could apply voltage as a pulse of a given length from 1
to $10^5\;\mu$s. The pulse could be repeated with a frequency $f$. At
a prescribed moment of time a short flash of light $\geq 5\;\mu$s was
on and an optical microscope image of the electrodes was made during
the flash. The applied voltage and the resulting current through the
electrodes were recorded. Alternatively keeping one electrode
grounded we applied voltage pulses of alternating polarity to the
working electrode repeated with some frequency. In this case bubbles
containing both gases H$_2$ and O$_2$ could be formed \cite{Sve11}.

The voltage needed to start the electrolysis was always larger than
that for macrosystems, $U_0=2.5-4\;$V. The high threshold voltage was
already observed in microsystems \cite{Hue02,Ate04,Men08}, but the
reason was not explained. It has to be stressed that in the
microsystems a planar geometry is used when both electrodes are in
the same plane. In this case the ohmic losses (resistance between the
electrodes) are considerable even for well conducting electrolytes.
This is in contrast with the macrosystems, where these losses are
usually negligible. For one pair of electrodes we measured the
resistance to be $R=2.3\;$k$\Omega$ but it varies to some degree from
pair to pair due to different geometry. The working current in our
system $I\sim 1\;$mA was quite large. Then the ohmic polarization is
estimated as a few volts ($2-3\;$V).

Conventionally we defined the threshold for the electrolysis applying
different voltages for $5\;$ms and observing the gas formation. At
$U=U_0$ the first gas was visible in the system. For example, for the
same pair of electrodes, for which the resistance was measured, we
found $U_0=3.2\;$V. In some cases $U_0$ was as large as $6\;$V or so,
but we have found out that this was due to the potential drop on the
electrode surface. Treatment of such chips in O$_2$-plasma
significantly reduced $U_0$. Probably a film (hydrocarbons) was
formed on top of the electrodes during the fabrication process. It
should be stressed that the absolute values of the potential has to
be taken with care because we deal with the two-electrode system and
there is uncertainty in the potential of the working electrode in
contrast with the three-electrode system.

\subsection{Current}\label{curr}

\begin{figure}[tb]
  \includegraphics[width=8cm]{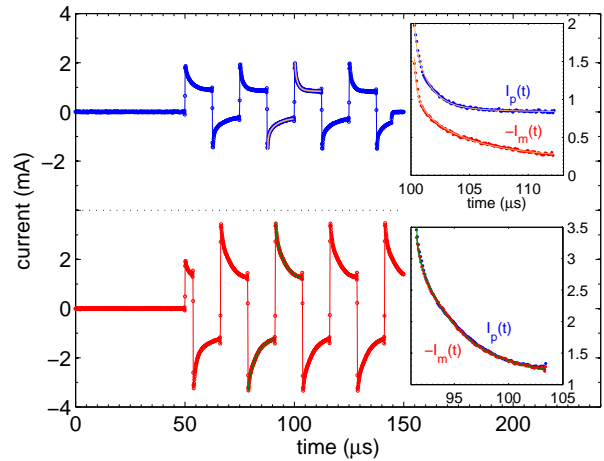}
  \caption{(top) Current response to the positive square voltage pulses
  of the amplitude $U=5.2\;$V and frequency $f=40\;$kHz. (bottom) Response
  to the voltage pulses of alternating polarity $U=6\;$V, $f=40\;$kHz.
  The insets show the currents from the main graph for half of one period
  and the fits of the currents with Eq. \ref{I_fit}.
  }\label{fig2}
  \vspace{-0.5cm}
\end{figure}

The current response to square voltage pulses applied to the working
electrode is shown Fig. \ref{fig2}. The response is in accordance
with the expectations. One can separate two components in the
current. The first one is the faradaic current $I_F$ that does not
depend on time $t$ for square voltage pulses. The second component is
responsible for the charging-discharging of the double layer on the
electrode surface and is not related to the electrochemical reaction.
It behaves as $I_1 e^{-t/\tau_c}$, where $I_1$ is a constant and
$\tau_c$ is the relaxation time related to the capacitance of the
double layer. As the result we fit the current response on the square
voltage pulse by a function of three parameters:
\begin{equation}\label{I_fit}
    I(t)=I_F+I_1 e^{-t/\tau_c}.
\end{equation}
It has to be stressed that typically in macro and microsystems
considered on the long-time scale the faradaic current $I_F$ depends
on the time because it is diffusion limited. For parallel electrodes
the current is proportional to $\sqrt{\tau_D/t}$ in accordance with
the Cottrell equation \cite{Bar80}, where $\tau_D$ is the diffusion
time for the electroactive species in the solution. In our case the
diffusion time is too long to play any role. Roughly it can be
estimated as $\tau_D\sim L^2/D_a\sim 100\;$ms, where $L\sim 10\;\mu$m
is a typical size of our electrodes and $D_a\sim 10^{-9}\;$m$^2$/s is
a typical diffusion coefficient of the electroactive species in
water.

We used Eq. (\ref{I_fit}) to fit the current response as shown in
Fig. \ref{fig2}. The quality of the fit can be seen in the insets.
The faradaic current can be well approximated by a constant in the
investigated time domain $t < 100\;\mu$s. For alternating pulses the
positive $I_p(t)$ and negative $I_m(t)$ halves of the pulse are
symmetric as shown in the bottom inset. For single polarity pulses
the relaxation time $\tau_c$  is larger for the negative half of the
pulse (4.9 against $1.3\;\mu$s) reflecting dependence of the double
layer capacitance on the potential \cite{Bar80}. For alternating
polarity pulses in Fig. \ref{fig2} we have found from the fit
$\tau_c=4.0\;\mu$s. With the frequency increase the relaxation time
decreases slowly indicating that not all adsorption states can be
filled for a short time. It has to be noted that the Faraday
component of the current persists up to the shortest pulses of
$1\;\mu$s ($f=500\;$kHz), which we investigated. Presence of this
component signals formation of gases on this shortest time scale that
was directly observed with the stroboscope.

The faradaic current reaches rather large values. For the case shown
in Fig. \ref{fig2} it is $I_F=0.80\pm 0.07\;$mA for the single
polarity pulses and $I_F=1.08\pm 0.03\;$mA for the alternating
polarity voltage pulses. Keeping in mind that the open area of the
electrode was around $1260\;\mu$m$^2$ one finds that the current
density was $j_F\sim 80\;$A/cm$^2$. This is the average value of the
current density but locally it can be 5 times larger (see Sec.
\ref{estimates}). This value is much larger than the maximal value
$\sim 1\;$A/cm$^2$ \cite{Vog05} that can be reached in the long-time
electrolysis.

\bigskip

The most important conclusions that can be drawn from the
current-voltage behavior of the microsystem on the short-time scale
are the following. (i) The electrolysis persists normally up to the
shortest investigated time $1\;\mu$s. (ii) The current density is at
least two orders of magnitude larger than was previously observed in
the macrosystems.

\section{Stroboscopic observations}\label{strob}

The stroboscopic technique gives images that were made for different
runs. To make conclusions from these images the system has to behave
reproducibly from run to run. For our system this condition was
fulfilled. The images made at the same moment for different runs were
hardly distinguishable on the short-time scale and varied slightly on
the long-time scale $>1\;$ms. In the latter case the positions and
sizes of separate bubbles were slightly different but qualitatively
the pictures were the same. For good visibility we used  typically a
$10\;\mu$s long flash of green light that was delayed for different
times from the beginning of the voltage pulse. The camera collected
the light from the microscope during the flash. The estimated time
uncertainty in 5 $\mu$s is reasonably small although comparable with
the time scale. The gas produced by series of pulses in most cases
disappeared diffusively in less than 1 s. Sometimes when the amount
of produced gas was large a stable (pinned) bubble formed. In this
case we have to refill the chip for further use. Due to evaporation
we have been able to operate with a filled chip during 15 min.

\begin{figure}[tb]
  \includegraphics[width=8cm]{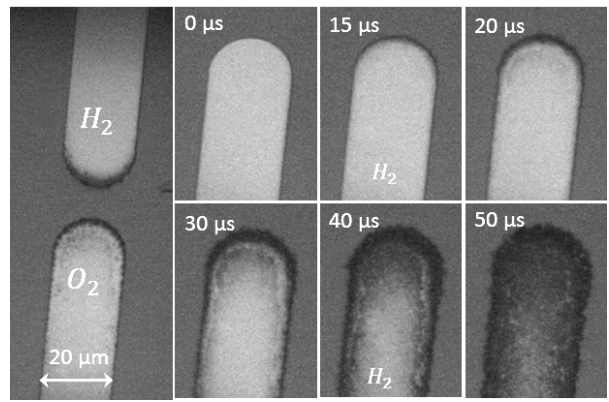}
  \caption{(left panel) Positive potential is applied to the working
  electrode (bottom). The photo was made $20\;\mu$s after switching on
  the voltage. Six images on the right show the working electrode
  in different moments of time when negative voltage is applied to it.
  The indicated time is taken in the middle of the flash.
  }\label{fig3}
  \vspace{-0.5cm}
\end{figure}

Figure \ref{fig3} shows the electrodes in different moments of time.
A positive or negative voltage pulse of $50\;\mu$s long was applied
to the electrodes. The left image shows both electrodes $20\;\mu$s
after beginning of the positive voltage pulse. Oxygen and hydrogen
are formed above the working electrode and grounded electrode,
respectively. All the other images show the working electrode with
hydrogen in different moments during the negative voltage pulse.

Indeed, the periphery of the electrode is a preferable place for
bubble nucleation, but one can see some haze in the inner part of the
electrode.  With our microscope we can reliably resolve bubbles with
a diameter of $500\;$nm. The bubbles forming the haze are smaller. We
are able to see them because they strongly scatter the light. The
diameter of these scattering centers is estimated as
$d>\lambda/\pi\approx 170\;$nm \cite{Boh83} where $\lambda=520\;$nm
is the wavelength of the used LED. These scattering centers fill
densely the inner part of the electrode and grow with time.

\subsection{Supersaturation}\label{estimates}

Let us make simple estimates. For the time $\tau=20\;\mu$s the gas
molecules can diffuse in the vertical direction off the electrode on
the distance
\begin{equation}\label{dist}
    l_{H_2}\sim\sqrt{D_{H_2}\tau}\approx 300\;\textrm{nm},\ \ \
    l_{O_2}\approx 200\;\textrm{nm},
\end{equation}
where $D_{H_2}=4.5\times 10^{-9}\;$m$^2$/s and $D_{O_2}=2.0\times
10^{-9}\;$m$^2$/s are the diffusion coefficients of hydrogen and
oxygen in water. These values of the diffusion length have to be
considered as the upper limits on the bubbles that can be formed
during the time $\tau$. Therefore, it is natural that we can see only
nanobubbles during the first $20\;\mu$s of electrolysis.

The relative supersaturation for H$_2$ or O$_2$ molecules can be
found as follows. The concentration of $i$-th gas in the diffusion
layer is $n_i={\cal N}_i/l_iA$, where $A$ is the area of the working
electrode and ${\cal N}_i$ is the number of molecules produced by the
electrolysis. This number is related to the Faraday current as ${\cal
N}_{H_2}=I_F\tau/2|e|$ (for H$_2$), where $e$ is the electron charge.
The relative supersaturation is the ratio of the gas concentration to
the saturated concentration $n^{(s)}_i$, which at normal conditions
is $4.7\times 10^{17}\;\textrm{cm}^{-3}$ or $7.7\times
10^{17}\;\textrm{cm}^{-3}$ for hydrogen or oxygen, respectively. Then
we find for the supersaturation
\begin{equation}\label{SS}
    S_{H_2}=\frac{j_F}{2|e|n_{H_2}^{(s)}}l_{H_2},
    \ \ \
    S_{O_2}=\frac{j_F}{4|e|n_{O_2}^{(s)}}l_{O_2},
\end{equation}
where $j_F=I_F/A$ is the current density.

The faradaic current recorded together with the images in Fig.
\ref{fig3} was $I_F\approx 1.5\;$mA. However, it is not
straightforward to estimate the current density because it is highly
nonhomogeneous for the planar electrodes. We already analyzed the
current distribution over the electrode (see \cite{Sve11}
Supplemental) extracting the information from the observed "wear" of
the electrodes. It was found that the current density over the
electrode surface can be presented as $j_F(x,y)=j_{m}f(x,y)$, where
$x$ and $y$ are the coordinates of a point on the electrode and
$j_{m}$ is the maximal current density. The function $f(x,y)$ has its
average value over the electrode $\overline{f(x,y)}\approx 0.2$. The
area of the electrode was $A\approx 1260\;\mu$m$^2$ and we find
$j_{m}\approx 600\;$A/cm$^2$. This is an extremely high value in
comparison with usual electrolysis and it is realized in the area of
active bubble formation as can be seen in Fig. \ref{fig3}. The
supersaturation in this area for the time $\tau=20\;\mu$s is
estimated as
\begin{equation}\label{SS1}
    S_{H_2}^{m}\approx 2700,\ \ \ S_{O_2}^{m}\approx 1200.
\end{equation}
These are incredibly large values.

\subsection{Homogeneous bubbling}\label{homogen}

Observation of the haze far from the electrode periphery is a signal
of homogeneous nucleation of small bubbles. Our estimates of the
supersaturation (\ref{SS1}) also support this idea.  In the classical
theory of homogeneous nucleation \cite{Deb96,LL5} the probability to
create a bubble of the critical size $r_c$ is given by the exponent
\begin{equation}\label{prob}
    w\sim \exp\left\{-\frac{4\pi r_c^2\gamma}{3k_BT} \right\},\ \ \
    r_c=\frac{2\gamma}{\Delta P},
\end{equation}
where $\gamma$ is the surface tension of the liquid and $\Delta P$ is
the pressure difference between gas and liquid. The pressure in gas
$P_g$ must support the high supersaturation in liquid. It is defined
as $P_g = SP_s$, where $P_s$ is the saturated pressure for a gas at
normal conditions. For both H$_2$ and O$_2$ gases $P_s\approx 1\;$atm
and we find $\Delta P_i\approx P_a (S_i-1)$ for the $i$-th gas.

For small supersaturation $S-1\sim 1$ the critical size $r_c\sim
1\;\mu$m is large and the exponent in (\ref{prob}) is incredibly
small. For high supersaturation $S\sim 1000$ the critical bubble
$r_c\sim 1\;$nm is small and the probability to create a critical
bubble, which is able to grow, becomes appreciable. It is hardly
reasonable to make more detailed estimates. This is because a number
of important corrections have to be included in the classical
nucleation theory \cite{Deb96,Gir90,Ani03}. It has to be noted that
according to (\ref{SS}) $S_i$ increases with time as $\sqrt{\tau}$.
Therefore, starting from some moment after beginning of the
electrolysis homogeneous bubble formation is expected. With the
temperature increase as one can see from (\ref{prob}) the homogeneous
bubbling must start earlier and due to the exponential dependence
this tendency must be well pronounced. This prediction is very easy
to check experimentally doing water electrolysis at elevated
temperatures.

\begin{figure}[tb]
  \includegraphics[width=8cm]{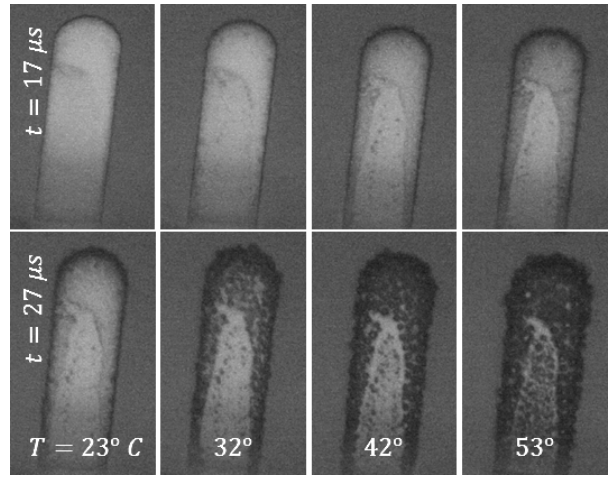}
  \caption{Images of the electrode at $t=17\;\mu$s (top row)
  and at $t=27\;\mu$s (bottom row) taken at different temperatures.
  A negative voltage pulse of $20\;\mu$s long was the
  applied to the electrode. Hydrogen nano and microbubbles are visible.
  }\label{fig4}
  \vspace{-0.5cm}
\end{figure}

For this experiment we heated the chip by a flat resistor and
controlled the temperature by a thermocouple. A $20\;\mu$s long
negative voltage pulse was applied to the working electrode. Images
shown in Fig. \ref{fig4} were taken at the moments $t=17\;\mu$s (top
row) just before the pulse end and at $t=27\;\mu$s (bottom row)
$7\;\mu$s after the end of the pulse. Some chemical remnants are
clearly visible on the electrode surface. They appeared due to
imperfections in the fabrication process and visualize the area with
the reduced ability for gas production. The temperature increase
resulted in the increase of the Faraday current from $I_F\approx
1.3\;$mA at $T=23^{\circ}\;$C up to $I_F\approx 2.5\;$mA at
$T=53^{\circ}\;$C. This increase is expected due to the increase of
the ions diffusion coefficients. However, the increase of the bubble
visibility and speed up of the dynamics are much more prominent.

Increase of the solution conductivity (or equivalently current) with
temperature was checked independently by measuring the conductivity
of the bulk solution with Mettler Toledo SevenMulti conductivity
meter. We have found that the conductivity can be well described by
the linear behavior $\sigma(T)=\sigma_0\left[1+\alpha(T-T_0)\right]$,
where $\sigma_0$ is the conductivity at temperature $T_0$. The
thermal coefficient was found to be $\alpha=0.024\;$K$^{-1}$. This
value is in a good agreement with the value $0.028\;$K$^{-1}$ found
from the current increase on the chip and with the values reported in
different literature sources.

Comparing the images in the top row of Fig. \ref{fig4} one can see
that much more nanobubbles becomes visible with the temperature
increase. On the other hand, the gas volume at $T=53^{\circ}\;$C must
be approximately twice larger than at $T=23^{\circ}\;$C as follows
from the current increase. It has the following explanation. At
higher temperature the nucleation of critical bubbles happened
earlier and to the moment $t=17\;\mu$s much more bubbles had enough
time to grow to a visible size. At $T=23^{\circ}\;$C most of the
bubbles are small enough to be invisible.

It becomes even more clear when one compares the top and bottom rows.
For example, the bottom image at $23^{\circ}\;$C contains much more
visible gas than the top image. However, when the pulse is over there
is no gas production anymore and one would expect the equal or the
opposite relation between the gas volumes. There is a very simple and
natural explanation for this phenomenon. The voltage pulse produces a
certain amount of gas. This gas can exist in three forms: molecules
dissolved in the liquid, nanobubbles larger than the critical size
but smaller than needed to be visible with the stroboscope, and
visible nano or microbubbles. If homogeneous nucleation of bubbles
happens, then at a certain moment a significant volume of gas is
collected in the invisible nanobubbles. These nanobubbles grow with
time and become visible with a delay producing the phenomenon of gas
appearing from nothing.

We consider Fig. \ref{fig4} and explanations above as a strong
support of the idea that in the short-time electrolysis we observe
homogeneous formation of bubbles.

\subsection{Transition to long-time dynamics}\label{long-time}

\begin{figure}[tb]
  \includegraphics[width=8cm]{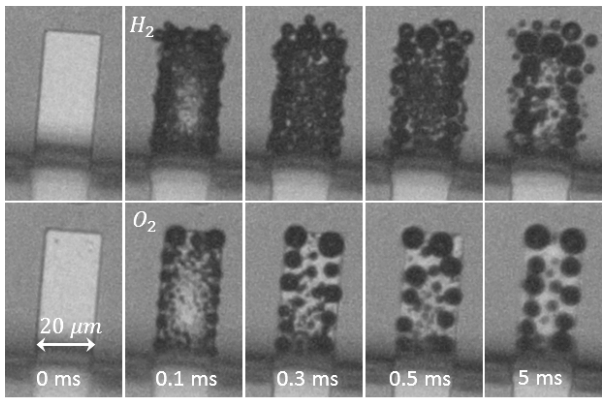}
  \caption{Long-time dynamics of bubbles. Hydrogen (top row) or
  oxygen (bottom row) bubbles were produced by a voltage pulse of
  $100\;\mu$s long. The images in each row were taken at different
  moments of time.
  }\label{fig5}
  \vspace{-0.5cm}
\end{figure}

Let us consider now how transition to the usual long-time
electrolysis happens. A negative or positive voltage pulse of
$100\;\mu$s long was applied to the working electrode. The electrode
was observed in different moments of time when the pulse was already
switched off. Some results are shown in Fig. \ref{fig5}.

It was already clear from Fig. \ref{fig4} that nanobubbles growing
from nucleus start to coalesce in some moment of time and form
microbubbles (see bottom row in Fig. \ref{fig4}). Fast aggregation
becomes possible due to high density of nanobubbles. Now from Fig.
\ref{fig5} one can see that significant changes happen between the
images taken at 100 and $300\;\mu$s. The following dynamics becomes
very slow. This is especially clear for the images at $0.5\;$ms and
$5\;$ms. It is expected that the long-time dynamics is controlled by
the slow diffusion and slow coalescence.

\bigskip

The main conclusion that one can make from the short-time
stroboscopic observations is that the nanobubbles are densely
produced in the homogeneous nucleation regime. They exist for some
time in the "invisible" form when they are small enough to scatter
the light and are invisible for the stroboscope. At later moments
they become visible but exist as separate nanobubbles. Later they
start to aggregate to form microbubbles that are usually observed in
the electrolysis. Extremely high current densities and relative
supersaturations are observed.

\section{Alternating sign pulses}\label{alternating}

Up to now we discussed only hydrogen and oxygen bubbles that are
formed on different electrodes when negative or positive voltage is
applied to the working electrode. However, there is a simple but
effective procedure that allows formation of bubbles containing
mixture of H$_2$ and O$_2$. Moreover, the gas composition can be
controlled electrically. The idea is to use pulses of alternating
polarity \cite{Sve11}. In this case hydrogen and oxygen will be
produced locally above the same spot on the electrode. If the pulses
are short enough to neglect the gas redistribution due to diffusion,
then one can expect formation of bubbles containing both gases.

For homogeneous nucleation formation of a bubble containing both
gases is favorable in comparison with two bubbles containing
different gases. This is because even in the regime of homogeneous
nucleation there is an energetic barrier to form the critical nuclei.
For the second gas it is easier to diffuse inside of the existing
bubble than to overcome the barrier and form a new nuclei.

\subsection{Reaction inside of nanobubbles}\label{reaction}

\begin{figure}[tb]
  \includegraphics[width=8cm]{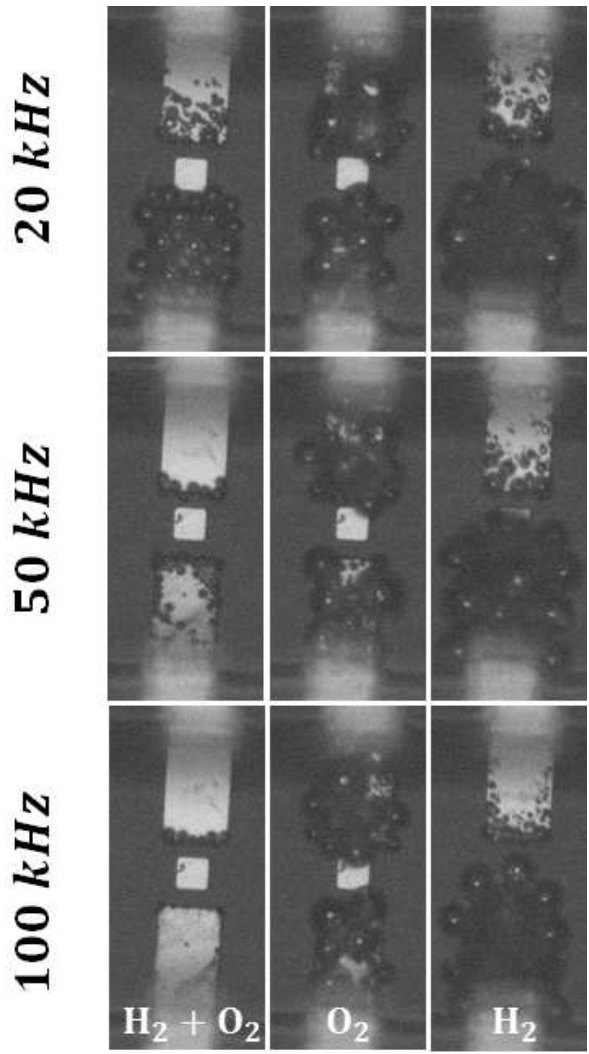}
  \caption{Gas production with the negative pulses (right column),
  the positive pulses (middle column), and the alternating polarity pulses
  (left column) at different frequencies. For the alternating pulses
  the visible gas production disappears suddenly at higher
  frequencies.
  }\label{fig6}
  \vspace{-0.5cm}
\end{figure}

While the frequency of alternating pulses is not very high the
process proceeds similar to that for single polarity pulses. However,
when frequency is higher than $f=20\;$kHz (at room temperature) the
visible gas suddenly disappears. This can be clearly seen in Fig.
\ref{fig6}. The figure shows gas developed in the system after
$1\;$ms of electrolysis. Negative potential pulses applied to the
working electrode (bottom in each image) result in H$_2$ formation
above this electrode at different frequencies (right column). No
significant dependence on the frequency is observed. Oxygen formed
above the working electrode for positive pulses also does not show
frequency dependence (middle column). The left column shows the
situation for the alternating pulses applied to the working
electrode. In this case we observe strong dependence on frequency.
The alternating pulses repeated with the frequency $f=20\;$kHz
produce significant amount of gas. One can see that this gas is
collected in bubbles which differ from that in the middle and right
columns. Already at $f=50\;$kHz only small volume of gas is observed.
At $f=100\;$kHz practically no gas is visible.

\begin{figure}[t]
    \vspace{0.5cm}
  \includegraphics[width=8cm]{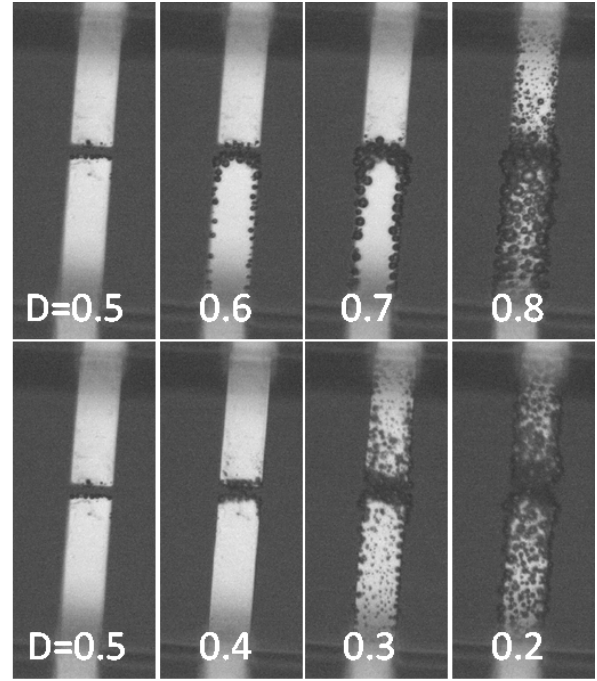}
  \caption{Visible gas in the system after $200\;\mu$s of processing
  at $f=100\;$kHz for different duty cycles $D$ of alternating polarity
  pulses. When the working electrode is equal
  time at positive and negative potential ($D=0.5$) no gas is visible.
  If each pulse is more negative ($D>0.5$), then excessive hydrogen is
  formed. For $D<0.5$ excessive oxygen appears. In all cases the deviation
  of $D$ from 0.5 results in reappearance of the visible gas.
  }\label{fig7}
  \vspace{-0.5cm}
\end{figure}

Where the gas has disappeared? The Faraday component of the current
depends on frequency very weakly if depends at all. Because the
pulses of single polarity produce the gas at all frequencies it is
difficult to assume that the Faraday's law breaks at higher
frequencies. Experimentally we established that the gas disappearance
is related to the stoichiometric production of hydrogen and oxygen.
There is no visible gas production while positive and negative halves
of the pulse have the same duration. In this case two H$_2$ molecules
are produced for one O$_2$ molecule. If we change the relative
duration of the positive and negative parts of the pulse then the
stoichiometric balance will be broken. In this case the gas
reappeared in the system as shown in Fig. \ref{fig7}. In this figure
$D$ is the duty cycle of the pulse that is the fraction of time when
voltage on the working electrode is negative.

It seems reasonable to assume that the gas disappears due to a
chemical reaction between gases happening in the solution. The
reaction is hardly possible between separate dissolved gas molecules.
For high supersaturation the molecules exist in the solution only a
short time before entering nanobubbles as one can see in Fig.
\ref{fig3} and \ref{fig4}. It is possible that the reaction proceeds
in nanobubbles containing stoichiometric mixture of H$_2$ and O$_2$.
Of course, the bulk mixture of gases at room temperature and normal
pressure will not react, but the gases confined in a small volume
could behave differently. It is known that below a certain size,
material properties can change drastically \cite{Hod07}. For example,
the surface tension can support metastable phases of nanocrystals
\cite{Mei07} that exist only at high pressure for bulk materials. In
liquids the surface tension results in significant pressure inside of
nanobubbles. This pressure can shift the chemical equilibrium
\cite{San00} and it is known also that fast dynamical processes play
role for ignition of the reaction on the macroscale
\cite{Tro92,Gol07,Dry07}. Nevertheless, the precise mechanism of the
reaction is not clear at this moment.

\begin{figure}[tb]
    \vspace{0.5cm}
  \includegraphics[width=8cm]{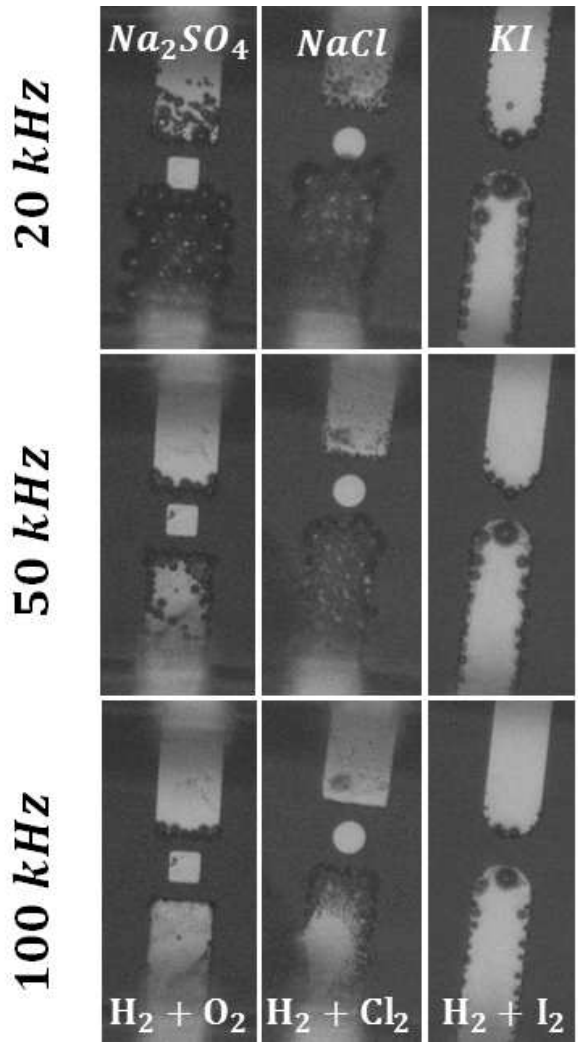}
  \caption{Alternating pulse electrolysis in different solutions:
  Na$_2$SO$_4$, NaCl, and KI dissolved in water. Each column corresponds
  to one of the solutions labeled on the top. Gases formed in the electrolysis are
  indicated in the bottom of the columns. Each row corresponds to
  the same frequency. The effect of gas disappearance clearly
  visible for Na$_2$SO$_4$ solution becomes less pronounced for NaCl
  and disappears at all for KI solution.
  }\label{fig8}
  \vspace{-0.5cm}
\end{figure}

We investigated also the effect of gas disappearance in different
solutions such as Na$_2$SO$_4$, NaCl, and KI dissolved in deionized
water to similar concentrations ($\approx 1\;$M). The results of
alternating pulse electrolysis in these solutions are shown Fig.
\ref{fig8}. In the case of NaCl hydrogen and chlorine are produced.
These gases can react with each other in the exothermic reaction with
the enthalpy $\Delta H=-92\;$kJ/mol considerably smaller than that
for hydrogen and oxygen $\Delta H=-242\;$kJ/mol. One can see that the
amount of visible gas decreases with the frequency increase but does
not disappears completely. It means that the reaction starts
spontaneously in smaller bubbles than for Na$_2$SO$_4$ solution. In
the case of KI solution hydrogen and iodine are produced. The
reaction between these molecules is endothermic, $\Delta
H=25\;$kJ/mol, and cannot proceed spontaneously. Only weak dependence
on $f$ is observed that is rather related to a small variation of the
current.

Below we provide some additional experimental information that
supports the hypothesis of chemical reaction inside of nanobubbles.
However, the reaction is not the main topic of this paper; more
information can be found in \cite{Sve11}.

\subsection{Observations with vibrometer}\label{vibrometer}

Observation of gas with a vibrometer is complimentary to observation
with the stroboscope. The stroboscopic picture becomes visible when
the bubbles significantly scatter the light. The nanobubbles with the
size $d<\lambda/\pi$ are invisible for the stroboscope. On the other
hand, strong scattering of the laser beam makes the interferometric
(vibrometer) measurements impossible. Therefore, the vibrometer is
effective for very small bubbles $d<\lambda/\pi$ where the
stroboscope fails.

The vibrometer is sensitive to variations of the optical path $\Delta
d(t)$ of the laser beam with time. The optical path is defined by the
change of the refractive index $\Delta n$ due to the presence of gas
in the solution. If we assume that within the laser beam the lateral
distribution of gas is homogeneous then
\begin{equation}\label{Dd}
    \Delta d=\int\limits_0^{\infty}\textrm{d}z\left[\Delta n_1(z)+
    \Delta n_2(z)\right],
\end{equation}
where $z$ is the vertical coordinate counting from the electrode
surface, and $\Delta n_i(z)$ is the change of the refraction index
due to gas $i$. Note that $\Delta d$ does not change if the gas
molecules redistribute along $z$ but their number stays the same.

The refractive index of a liquid containing dissolved gas or gas
collected in very small bubbles can be calculated using the Bruggman
effective medium approximation \cite{Asp82}. If the volume fraction
of gas is small, $f_i \ll 1$, then
\begin{equation}\label{Brug}
    \Delta n_i\approx\frac{3n_0(1-n_0^2)}{2(1+2n_0^2)}f_i,
\end{equation}
where $n_0\approx 1.34$ is the refractive index of the solution at
$\lambda = 633\;$nm. The volume fraction of gas is expressed as $f_i
= N_i/N_{sol}$ , where $N_i$ is the concentration of $i$-th gas and
$N_{sol}\approx 3.4\times 10^{22}\;$cm$^{-3}$ is the concentration of
the solution. If the gas in the volume cannot disappear, for example,
due to chemical reaction, then the vibrometer signal must be
proportional to the total gas flux produced by the Faraday current
\cite{Sve11}:
\begin{equation}\label{velocity}
    v(t)=-\frac{\partial\Delta d}{\partial t}\approx\frac{0.35}{N_{sol}}
    \left[J_1(t)+J_2(t)\right],
\end{equation}
where $v(t)$ is the signal of the vibrometer and its opposite sign is
defined by the instrument. The total flux of the produced molecules
$J_{tot}=J_1+J_2$  is a non-decreasing function of time.

\begin{figure}[tb]
    \vspace{0.5cm}
  \includegraphics[width=8cm]{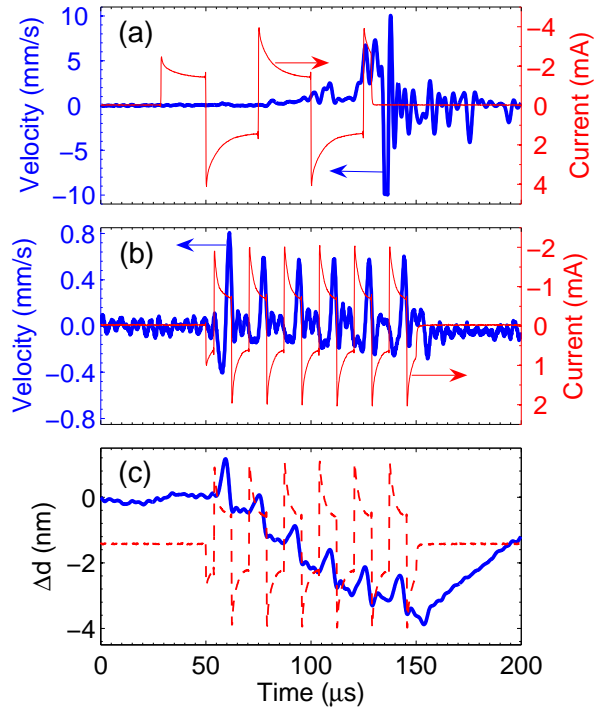}
  \caption{Vibrometer signal. (a) Processing at low frequency,
  $f=20\;$kHz, results in formation of large bubbles strongly
  scattering the light. The thick blue curve is the signal and the
  thin red curve is the current. (b) The signal and current as
  functions of time for $f=60\;$kHz. The peaks in the signal are
  in phase with the voltage (current) pulses. (c) The signal in (b)
  integrated over time (thick blue). The red dashed line presents the
  current and is given here only for eye guidance.
  }\label{fig9}
  \vspace{-0.5cm}
\end{figure}

We observed the process with an instrument Polytec MSA-400. The laser
beam ($\lambda=633\;$nm), with a diameter of $1.5\;\mu$m, was focused
on the electrode in its center at small distance from the edge. The
signal is presented in Fig. \ref{fig9}(a) and (b). When frequency of
alternating sign pulses was low, large bubbles can be formed that
scatter light significantly. In this case one cannot extract helpful
information from the signal as shown in Fig. \ref{fig9}(a) for
$f=20\;$kHz. However, for higher frequencies the bubbles small in
comparison with $\lambda/\pi$ are formed and one can clearly see the
signal (see Fig.\ref{fig9}(b)). The signal becomes weaker in the
inner areas of the electrode in accordance with the current
distribution. It disappears at all outside of the electrodes. The
latter demonstrates that the signal cannot be related to any kind of
vibrations. Integration of the signal over time gives the variation
of the optical path $\Delta d(t)$. The integrated signal is presented
in Fig. \ref{fig9}(c).

The function $\Delta d(t)$ demonstrates prominent maximums that are
in phase with the voltage pulses. These maximums mean periodic
decrease of the gas concentration in the liquid. The only reasonable
explanation for this decrease is the reaction between hydrogen and
oxygen. Of course, the vibrometer signal is sensitive only to the
overall concentration of gas in the liquid. It is not possible to say
if the gas interacts as dissolved molecules or inside of nanobubbles.
However, from observations of hydrogen and oxygen on the same time
scale we know that each gas  is collected in nanobubbles. It is
hardly possible that the gases produced by alternating polarity
pulses will behave differently.

The linear trend in $\Delta d(t)$ means that not all produced gas is
burned in the reaction but its small part is left after each period.
This residual gas exists mainly in the form of dissolved molecules.
When the gas production is switched off the residual gas disappears.
To all appearance it disappears also as the result of reaction
because reduction of gas concentration due to diffusion has to
proceed slower than observed. However, this point needs more
attention.

\subsection{Modification of the surface of electrodes}\label{wear}

\begin{figure}[tb]
    \vspace{0.5cm}
  \includegraphics[width=8cm]{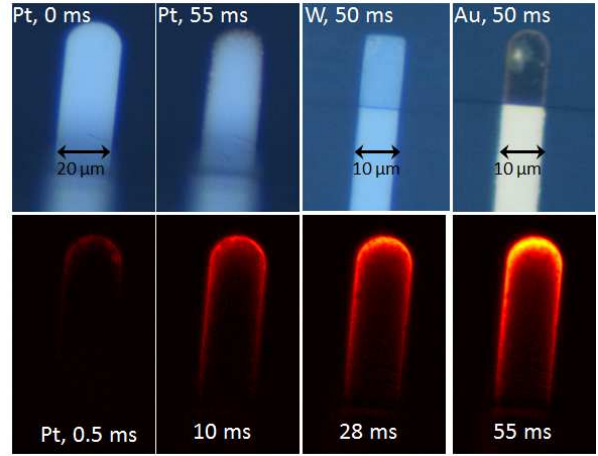}
  \caption{Modification of the electrode surface. (top row) The left
  image shows untreated Pt electrode. The second image (from left to
  right) is the same electrode after $55\;$ms of processing. The third
  and forth images are for W and Au electrodes, respectively, after $50\;$ms of
  processing. (bottom row) Successive images of changes in the Pt electrode for
  different processing times. The colormap is changed for better
  visibility.
  }\label{fig10}
  \vspace{-0.5cm}
\end{figure}

The effect of gas disappearance is accompanied by modification of the
electrode surface (see Fig. \ref{fig10}). We call this phenomenon
"wear"of electrodes. It never has been observed for single polarity
electrolysis but well visible for the alternating polarity
processing. The effect is observed for all investigated materials Pt,
Pd, W, and Au but manifests itself to a different degree. The
strongest effect is observed for gold and the weakest one is for
tungsten. It is in correlation with the material yield strength. The
effect increases with the process time as shown in the bottom row of
Fig. \ref{fig10}.

The chemical origin of the surface modification can be excluded
because it is observed for different metals. On the other hand, we
can exclude the electrochemical origin of the modification because it
is not observed for single polarity pulses. We expect that the
surface modification happens as the result of local release of energy
from the exploding nanobubbles that mechanically modify and shift the
material of electrodes. This assumption naturally explains
correlation with the material yield strength.

We observed only weak modification of the electrodes for the
alternating pulse electrolysis in the NaCl solution. In the KI
solution there was no visible modification of the electrodes at all.
These facts are in good agreement with the enthalpy of the reactions
between H$_2$ and O$_2$, Cl$_2$, or I$_2$: the larger the enthalpy
the stronger is the modification of the surface.

\begin{figure}[tb]
    \vspace{0.5cm}
  \includegraphics[width=8cm]{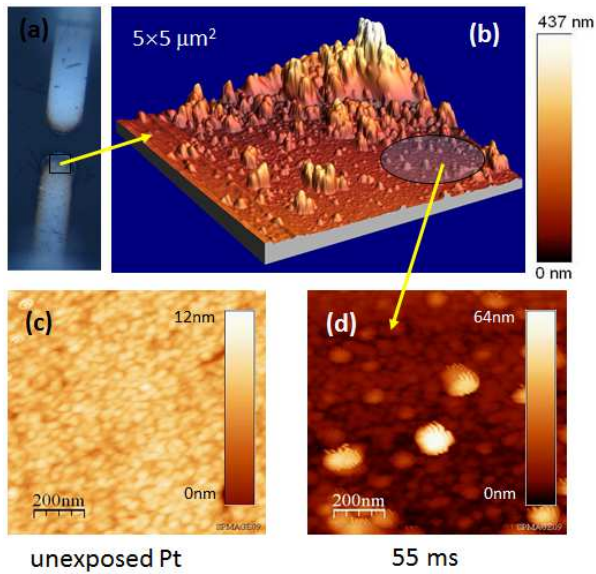}
  \caption{AFM scan of the modified surface. The process time was $55\;$
  ms (a) Optical image of the electrodes in air. An approximate location
  of the AFM scan is shown. (b) 3D topography of the electrode. Significant
  displacement of the material is observed. (c) Platinum surface before
  the process (normal roughness). (d) Modified roughness of the electrode
  after the process.
  The roughness changes on both lateral and  vertical scales.
  }\label{fig11}
  \vspace{-0.5cm}
\end{figure}

The assumption of the mechanical modification is supported by the
analysis of the modified surface with an atomic force microscope
(AFM). The AFM scan of the modified electrode is shown in Fig.
\ref{fig11}. One can see significant displacement of the material in
Fig. \ref{fig11}(b), but the most important is that the process
changes completely the roughness topography of the original platinum
film as one can see in Fig. \ref{fig11}(c) and (d).

\bigskip
The experimental information collected so far on the alternating
pulse electrolysis (see also \cite{Sve11}) provides strong evidence
that the bubbles containing stoichiometric mixture of H$_2$ and O$_2$
gases explode as the result of spontaneous reaction between gases if
the bubble size is smaller than $100-200\;$nm. For pulses of low
frequency the stoichiometric composition is reached for larger
bubbles, which survive and can be observed with the stroboscope.
Mechanism of the reaction in small bubbles is still not clear.

\section{Discussion}\label{discussion}

Let us summarize what we learned about the transient nanobubbles
using different methods of observation. Of course, any bubble growing
from a nuclei to its actual size gets through the phase when its size
is in the nanorange. In this paper we have been concentrated on the
bubbles that grow and live in the nanophase a very short time $\sim
10\;\mu$s. This regime is realized only when the bubbles are
homogeneously produced nearby the electrode surface.

\subsection{Evolution of transient nanobubbles}

Homogeneous bubble production is closely related to the high current
density that results in a huge supersaturation within the diffusion
layer. The observed current density is large because we consider the
system on the short-time scale $\tau\sim 10\;\mu$s. In this case the
current is not restricted by the diffusion of electroactive species
to the surface as it happens for the long-time electrolysis.

For single polarity pulses the critical nucleus are produced by the
density fluctuations that is possible because of very high
supersaturation. These nucleus start to grow diffusively. On the time
scale $\tau\sim 10\;\mu$s they grow to nanobubbles of the size
$d\lesssim 200\;$nm close to the limit of stroboscopic visibility. At
longer time the separate nanobubbles start to aggregate to form
microbubbles. Already at $\tau\sim 100\;\mu$s most of the bubbles
exist as microbubbles. Small increase of the temperature on $10-20$
degrees strongly intensify the bubble nucleation and grows as it
should be for homogeneous bubble production. At higher temperature
the microbubbles appeared already at $\tau\sim 30\;\mu$s or even
earlier.


When all the gas is collected in microbubbles the supersaturation in
the surrounding liquid is low and the process becomes very slow. The
microbubbles slowly dissolve diffusively. For example, a bubble of
$10 \;\mu$m in diameter will dissolve for a time of the order of
$100\;$ms.

The bubbles produced by the alternating polarity pulses evolve
differently. The first half of the pulse produces high
supersaturation of one gas, for example, H$_2$. Nucleus of hydrogen
are formed homogeneously and grown diffusively. The second half of
the pulse produces O$_2$ molecules in the same location above the
electrode, but for oxygen it is energetically preferable to diffuse
in the existing hydrogen bubble than to form a new O$_2$ bubble. This
is why the alternating polarity pulses produce bubbles containing
mixture of hydrogen and oxygen. This mixture is stoichiometric if
positive and negative halves of the pulse have the same duration.

If a hydrogen bubble has grown large enough before oxygen starts to
diffuse inside, then such a bubble evolves similarly to the bubbles
containing only one gas. The lower the frequency of the alternating
pulses the longer time has a bubble to grow. The critical frequency
is around $20\;$kHz. The critical size of the bubble cannot be
estimated very precisely from the experiment, but roughly it is
around $d\approx 150\;$nm \cite{Sve11}. For high frequency,
$f>20\;$kHz, a bubble has no time to grow large. The gas mixture in
the bubble becomes stoichiometric while its size is still under
$150\;$nm. It was established (see also \cite{Sve11}) that in such a
bubble hydrogen and oxygen react spontaneously with the bubble
disintegration due to large released energy.

The mechanism of the reaction inside of nanobubbles is not clear.
Bulk mixture of gases does not react at room temperature and the
pressure that is equivalent to the Laplace pressure
$P_c=4\gamma/d\approx 30\;$bar for $d=100\;$nm. Of course, we cannot
apply the classical combustion theory \cite{Lew87} to a strongly
confined system such as nanobubble. To all appearance fast
microsecond dynamics plays an important role for the combustion of
mixture of gases. The detailed mechanism of the reaction is an
important point that needs further attention.

The transient nanobubbles considered in this paper cannot help with
the understanding of stability of the surface or bulk nanobubbles.
The stabilization mechanism can work in these bubbles too but they
cease to exists due to external reasons such as aggregation or
combustion. The transient nanobubbles exist in a very thin diffusive
layer above the surface, but we cannot say for sure if they attached
to the surface or not. Relation between the transient and
surface/bulk nanobubbles is an additional point that has to be
understood.

\subsection{Possible applications}

\begin{figure}[tb]
    \vspace{0.5cm}
  \includegraphics[width=8cm]{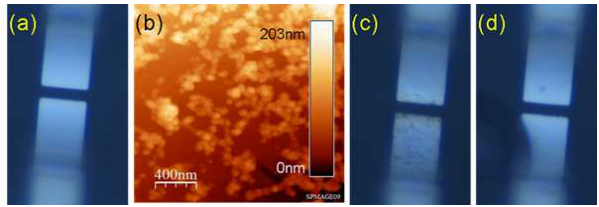}
  \caption{(a) Clean electrodes. (b) AFM image of the
  electrode surface after fouling with $50\;$nm nanoparticles.
  (c) Optical image of the fouled electrodes. (d) After
  cleaning with homogeneously produced oxygen nanobubbles.
  Similar effect is observed for hydrogen nanobubbles.
  }\label{fig12}
  \vspace{-0.5cm}
\end{figure}

We expect that the transient nanobubbles have a wide application
potential because of their unique properties. For example, dense
coverage with homogeneously produced nanobubbles can be used for
surface cleaning in electronic industry and other applications.
Figure \ref{fig12} demonstrates the proof of concept, where the
original clean electrodes were fouled with nanoparticles. The
following homogeneous production of nanobubbles cleans the surface.
The cleaning happens as the result of aggregation of nanobubbles that
pulls the nanoparticles out of the surface. The cleaning effect was
already stressed for the surface nanobubbles produced
electrochemically \cite{Wu_08} or with other methods
\cite{Liu08,Liu09,Yan11}. However, homogeneous bubble production is
better designed for local cleaning and has pronounced dragging effect
on nanoparticles.

Especially interesting applications are expected for the exploding
nanobubble phenomenon. Combustion inside of a nanobubble results in a
significant energy deposition that is localized in a nanovolume. It
can be applied, for example, to delete strongly adhered nanoparticles
from masks or wafers in the electronic industry. One can easy change
the bubble size by the electrical means (pulses frequency) varying
significantly the energy of explosion.

Combustion in microscopic volumes is considered as difficult or
impossible due to fast heat escape via the volume boundary
\cite{Lee02}. The situation becomes different for nanovolumes.
Exploding nanobubbles are the smallest combustion chamber realized in
nature \cite{Sma11}. It opens up a principal possibility to fabricate
an internal combustion engine microscopic in all three dimensions.

\section{Conclusions}

In this paper we described a special type of nanobubbles, which we
call transient nanobubbles. These bubbles exist very short time $<
100\;\mu$s before aggregating with other nanobubbles to form
microbubbles or before exploding disintegration due to reaction
between hydrogen and oxygen. The most distinctive feature for all
transient nanobubbles is their homogeneous production in the
electrochemical process. In contrast with the normal heterogeneous
bubbling these bubbles nucleate due to fluctuations of liquid
density.

Homogeneous nucleation is possible only for extremely high
supersaturations that is related to very high current densities.
These conditions can be met on a shot-time scale $\sim 10\;\mu$s when
the diffusion of charged particles in the electrolyte does not
restrict the current. On the short-time scale the gas produced on the
electrode surface has no time to diffuse far away from this surface.
For this reason high supersaturation can be realized only in a
diffusion layer of thickness $\sim 100\;$nm. In this thin layer only
nanobubbles can be formed.

Short voltage pulses of fixed polarity repeated with some frequency
$f$ can densely cover the electrode surface with nanobubbles
containing hydrogen or oxygen. These nanobubbles aggregate and form
microbubbles on the time scale $\sim 100\;\mu$s. The following
dynamics of microbubbles becomes very slow.

Short voltage pulses of alternating polarity produce nanobubbles
containing mixture of hydrogen and oxygen. If these bubbles are
smaller than $150\;$nm the gases inside of them react spontaneously.
As a result of this reaction significant energy is released in close
proximity of the electrode surface. Precise mechanism of the
spontaneous reaction inside of nanobubbles is still not clear and
needs clarification.

\acknowledgments We thank D. Lohse and J. Seddon for numerous
discussions and K. Ma, H. Dijkstra, M. Siekman, and H. Wolferen for
technical assistance. This work was partially supported by the Dutch
Technological Foundation (STW).


\end{document}